\documentclass[fleqn,10pt]{wlscirep}
\usepackage[utf8]{inputenc}
\usepackage[T1]{fontenc}

\title{Quantum imaging of a polarisation sensitive phase pattern with hyper-entangled photons}
\author[1]{Manpreet Kaur}
\author[2*]{Mandip Singh}
\affil[1]{Department of Physical Sciences, Indian Institute of Science Education and Research (IISER) Mohali, Sector-81, Mohali, 140306, India.}

\affil[*]{mandip@iisermohali.ac.in}

\begin{abstract}
 A transparent polarisation sensitive phase pattern makes a polarisation dependent transformation of quantum state of photons without absorbing them. Such an invisible pattern can be imaged with quantum entangled photons by making joint quantum measurements on photons.  This paper shows a long path experiment to quantum image a transparent polarisation sensitive phase pattern with hyper-entangled photon pairs involving momentum and polarisation degrees of freedom. In the imaging configuration,
a single photon interacts with the pattern while the other photon, which has never interacted with the pattern, is measured jointly in a chosen polarisation basis and in a quantum superposition basis of its position which is equivalent to measuring its momentum. Individual photons of each hyper-entangled pair cannot provide a complete image information. The image is constructed by measuring the polarisation state and position of the interacting photon corresponding to a measurement outcome of the non-interacting photon. This paper presents a detailed concept, theory and free space long path experiments on quantum imaging of polarisation sensitive phase patterns.
\end{abstract}
\begin{document}

\flushbottom
\maketitle
%
%
\thispagestyle{empty}


\section*{Introduction}
A transparent polarisation sensitive phase pattern introduces a position-dependent change in the phase of transmitted light depending on the polarisation of incident light. A classical image of such a pattern can be obtained by measuring the position-wise polarisation shift of the transmitted classical electromagnetic field in the image plane by a method known as polarisation contrast imaging \cite{pci1, pci2, pci3, faradayimaging, br_oct, brphase1, brphase2}. However, an image formed by a single photon exposure of the pattern in each execution of the experiment is defined as a quantum image. In this way, one can gain the quantum mechanical advantage over the classical imaging. One such advantage is the quantum secure transfer of images. Furthermore, by incorporating quantum entangled photons, one can utilise properties of quantum entanglement and quantum measurements to construct an image of the pattern.
Where a single photon of an entangled pair of photons interacts with the pattern. However, in contrast to the classical polarisation contrast imaging, complete information of the pattern is shared non-locally by both photons as a consequence of quantum entanglement, even if they are separated by a large distance when a photon interacted with the pattern. Here, one cannot obtain complete pattern information just by measuring a single photon. Therefore, a joint measurement becomes necessary and a quantum image of the pattern can be constructed by correlating the measurement outcomes. Because of quantum entanglement, quantum images can be transferred to another location securely and directly.

This paper presents the first free space long path experiment of quantum imaging of a transparent polarisation sensitive phase pattern with hyper-entangled photons. Where a hyper-entangled state is a product of quantum entangled states involving different degrees of freedom \cite{kwiat_hyp1, kwiat_hyp2}. In the experiment, the hyper-entangled state consists of a product of polarisation  entanglement \cite{chsh} and momentum entanglement of two photons \cite{zeirev1, horne3, shimony, mandelrev, eprboyd, dds_m}.  Where the momentum entanglement corresponds to Einstein-Podolsky-Rosen type of quantum entangled state originating from a finite region. Each photon of a hyper-entangled bi-photon state carries information of polarisation and momentum of the other photon but an individual photon has no well-defined position, momentum and polarisation.

A non-birefringent transparent phase object can be classically imaged with well-known phase-contrast imaging methods where the phase information is converted to intensity \cite{pc1,pc2,pc3}. In the context of the development of imaging experiments, a phase-contrast imaging method is applied for a non-destructive detection of a Bose-Einstein condensate \cite{nondesbec1, nondesbec2}. Two-photon coincidence quantum imaging of absorptive objects has been studied theoretically and experimentally with spatially correlated photons from a perspective of foundations of quantum mechanics \cite{gimage1, gimage3, gimage5, ghost4, ghim, qimaging, bar, shapiro2, raza,duality}. Quantum imaging of an absorptive object has been realised experimentally without detecting a photon interacting with the object \cite{zeilinger_1, zeilinger_2}. Bell's inequality violation experiments are performed with images \cite{bell_i, padgett_rev}. Two-atom ghost imaging of an absorptive pattern for atoms has been experimentally realised with metastable helium atoms \cite{gitruscot1, gitruscot2}.
 A polarisation-sensitive metasurface is imaged in the near field with polarisation entangled photons \cite{br_pol}.

In this paper, experiments involve a multi-dimensional entanglement in the form of a hyper-entangled state, to quantum image a transparent polarisation sensitive phase pattern in free space, where the pattern is positioned at a distance 16.91~m from a coincidence imaging camera. However, this particular experiment is not aiming to close loopholes. In the experiment, polarisation as well as momentum degrees of freedom of the hyper entangled photons are utilised for quantum imaging.

\section{Hyper-entangled photons}

\begin{figure*}
\centering
\includegraphics[scale=1.4]{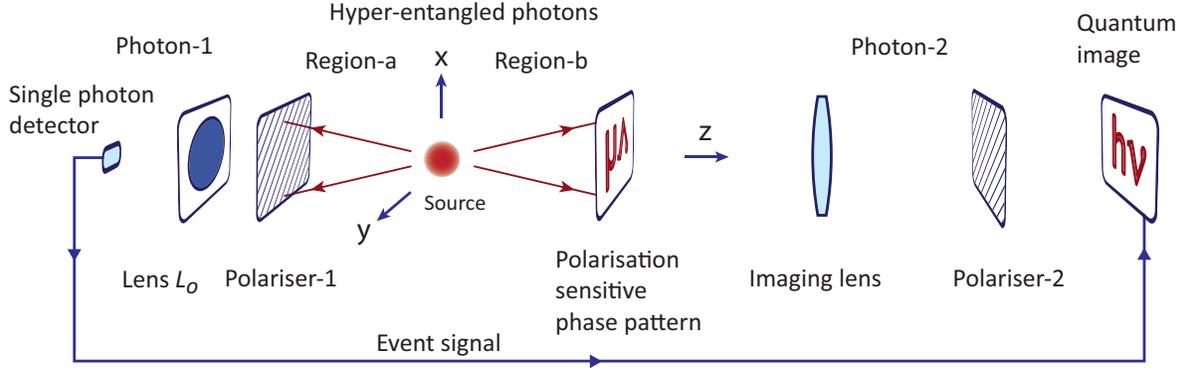}
\caption{\label{fig1} \emph{A schematic diagram of a quantum imaging experiment of a transparent polarisation sensitive phase pattern. A single photon-2 of a hyper-entangled photon pair is passed through the pattern. Polarisation state of each photon is measured by passing through polarisers. Photon-$1$ is measured in a particular quantum superposition of its position after its polarisation selection by passing through a lens $L_{o}$ followed by a detection by a single photon detector. An event signal produced by the single photon detector is sent to a single photon sensitive camera to detect a position of photon-2 corresponding to a particular measurement outcome of photon-$1$. An image is gradually formed by accumulating photon-$2$ detections on the camera by repeating the same experiment.
}}
\end{figure*}

Consider a hyper-entangled photons pair emitted by a source in opposite directions as shown in a schematic diagram Fig.~\ref{fig1}. Where photon-$2$ is passed through a transparent polarisation sensitive phase pattern. Since photons are polarisation entangled and momentum entangled separately therefore, a polarisation sensitive phase pattern induces a position-dependent variation of the quantum entangled state of photons. Polarisation measurement outcomes of photon-$1$ and photon-2 are determined by a measurement setting (orientation of pass axis) of polariser-$1$ and polariser-$2$, respectively.  After passing through  polariser-$1$, photon-$1$ is focused by a lens $L_{o}$ on a single photon detector after passing through a narrow aperture. A lens $L_{o}$ permits detection in a quantum superposition basis of the position of photon-1.  A photon detection produces a pulse named an event signal, which is sent to a single photon detection camera to allow registration of photon-$2$ corresponding to a selected measurement outcome. A quantum image is gradually formed by accumulating the photon registrations on camera for each repetition of the experiment with the same measurement setting. In this way, this is a coincidence imaging where the individual photon has no well-defined momentum and polarisation however, each photon is carrying momentum and polarisation information of the other photon by quantum entanglement. Therefore, measurement outcomes of both photons are required to construct a quantum image.

Consider Einstein Rosen Podolsky state (EPR) \cite{epr} of two particles in a position space, $|\alpha\rangle=\int^{\infty}_{-\infty}|x\rangle_{1}|x+x_{o}\rangle_{2} \mathrm{d}x$, where subscripts $1$ and $2$ are the labels of particles. In this state, both particles are equally likely to exist at all points in the position space with a constant separation $x_{o}$. In this way, they are spatially entangled. The EPR state in a momentum space can be written as $|\alpha\rangle=\int^{\infty}_{-\infty}e^{i \frac{p x_{o}}{\hslash}}|p\rangle_{1}|-p\rangle_{2}\mathrm{d}p$, where both particles have opposite momenta and are equally likely to exist at all points in the momentum space such that they are momentum entangled. Where $\hslash=h/2\pi$ is the reduced Planck's constant. In this way, both the position and momentum of each particle are completely unknown. Consider a source of a finite extension producing a pair of photons such that both photons are originating from the same position in the source. This is a case, if $x_{o}=0$ in the one-dimensional EPR state and in this paper, it is extended to three dimensions by including the polarisation of photons.

  Quantum state of photons originating from a position vector  $\mathbf{r}'$, in time independent case, leads to a finite amplitude to find a photon at a point $o_{a}$ in  region-$a$ and another photon at a point $o_{b}$ in  region-$b$ as shown in Fig.~\ref{fig1}. Consider points $o_{a}$ and $o_{b}$ prior to any optical element from the source.   A joint amplitude for photons to be at these points in three-dimensions is  $\frac{e^{ip_{1}|\mathbf{r}_{a}-\mathbf{r}'|/\hslash}}{|\mathbf{r}_{a}-\mathbf{r}'|}\frac{e^{ip_{2}|\mathbf{r}_{b}-\mathbf{r}'|/\hslash}}{|\mathbf{r}_{b}-\mathbf{r}'|}$ \cite{horne1, horne2} where $p_{1}$ and $p_{2}$ are magnitudes of momentum of photon-$1$ and of photon-$2$, respectively. Position vectors
 of points $o_{a}$ and $o_{b}$ from origin are $\mathbf{r}_{a}$ and $\mathbf{r}_{b}$, respectively. Distances of $o_{a}$ and $o_{b}$ from $\mathbf{r}'$ are $|\mathbf{r}_{a}-\mathbf{r}'|$ and $|\mathbf{r}_{b}-\mathbf{r}'|$, respectively. Here, both photons are produced from a same location within the source.
Consider, polarisation state of photon-$1$ propagating in region-$a$ is  horizontal $|H\rangle_{a}$ and of photon-$2$ propagating in  region-$b$ is vertical $|V\rangle_{b}$. A total amplitude to find  photon-$1$ at $o_{a}$  and photon-2 at $o_{b}$ is a linear quantum superposition of amplitudes originating from all points within the source. Therefore, a combined two-photon quantum state can be written as
\begin{equation}\label{eq1}
   |\Psi\rangle_{12}=A_{o} \int^{\infty}_{-\infty}\int^{\infty}_{-\infty}\int^{\infty}_{-\infty} \psi(x',y',z') \frac{e^{ip_{1}|\mathbf{r}_{a}-\mathbf{r}'|/\hslash}}{|\mathbf{r}_{a}-\mathbf{r}'|}\frac{e^{ip_{2}|\mathbf{r}_{b}-\mathbf{r}'|/\hslash}}{|\mathbf{r}_{b}-\mathbf{r}'|} \mathrm{d}x'\mathrm{d}y'\mathrm{d}z' \otimes|H\rangle_a|V\rangle_b
\end{equation}
 where $A_{o}$ is a normalisation constant and  $\psi(x',y',z')$ is an amplitude of pair creation at a location $r'(x',y',z')$ that is considered to be $ \frac{e^{-\frac{x'^{2}}{2\sigma^2_{x}}}}{(2\pi)^{1/2} \sigma_{x}} \frac{e^{-\frac{y'^{2}}{2\sigma^2_{y}}}}{(2\pi)^{1/2} \sigma_{y}} \frac{e^{-\frac{z'^{2}}{2\sigma^2_{z}}}}{(2\pi)^{1/2} \sigma_{z}}$.

 For a source extension much smaller than distances of points $o_{a}$ and $o_{b}$ from origin, Eq.~\ref{eq1} is written as
\begin{equation}
  \label{eq3}
   |\Psi\rangle_{12}=\Phi_{12}(r_{a};r_{b})\otimes|H\rangle_a|V\rangle_b
\end{equation}
where $\Phi_{12}(r_{a};r_{b})$ is the amplitude to find photon-$1$ at $o_{a}$ and photon-$2$ at $o_{b}$, which is calculated by solving the integral in Eq.~\ref{eq1},
\begin{equation}\label{eq4}
 \Phi_{12}(r_{a};r_{b})= A_{o} \frac{e^{i(p_{1}r_{a}+p_{2}r_{b})/\hslash}}{r_{a} r_{b}} e^{-((p_{1x}+p_{2x})\sigma_{x})^2/2\hslash^{2}} e^{-((p_{1y}+p_{2y})\sigma_{y})^2/2 \hslash^{2}} e^{-((p_{1z}+p_{2z})\sigma_{z})^2/2\hslash^{2}}
\end{equation}
where $p_{1x}=p_{1}\sin\theta_{a} \cos\phi_{a}$, $p_{1y}=p_{1}\sin\theta_{a} \sin\phi_{a}$ and $p_{1z}=p_{1}\cos\theta_{a}$ are the $x$, $y$ and $z$ components of momentum of photon-$1$ in the spherical polar coordinate system. Similarly $p_{2x}=p_{2}\sin\theta_{b} \cos\phi_{b}$, $p_{2y}=p_{2}\sin\theta_{b} \sin\phi_{b}$ and $p_{2z}=p_{2}\cos\theta_{b}$ are the $x$, $y$ and $z$ components of momentum of photon-$2$.

Photons are momentum entangled due to inseparability of $\Phi_{12}(r_{a};r_{b})$. Assume that photons have same energy therefore, magnitudes of their momenta are equal, $p_{1}=p_{2}$. Since photons are identical particles and there exist different amplitudes that cannot be distinguished if a photon is detected at $o_{a}$ and another photon is detected at $o_{b}$. An amplitude to find photon-$1$ at $o_{a}$ and photon-2 at $o_{b}$ is indistinguishable from finding photon-1 at $o_{b}$ and photon-2 at $o_{a}$ after exchange of photons if their polarisation is not measured.  Therefore, due to the Bosonic nature of identical photons, their symmetric quantum state is written as $|\Psi\rangle= \frac{1}{\sqrt{2}}(|\Psi\rangle_{12}+|\Psi\rangle_{21})$, where $|\Psi\rangle_{21}$ is a quantum state Eq.~\ref{eq1} after the exchange of photons.

After the exchange of photons, Eq.~\ref{eq3}  becomes $|\Psi\rangle_{21}=\Phi_{21}(r_{a};r_{b})\otimes|V\rangle_a|H\rangle_b$. However, from Eq.~\ref{eq4}, $\Phi_{21}(r_{a};r_{b})=\Phi_{12}(r_{a};r_{b})$. Therefore, a combined quantum state becomes
\begin{equation}\label{eq5}
   |\Psi\rangle=\Phi_{12}(r_{a};r_{b})\otimes\frac{1}{\sqrt{2}} \left(|H\rangle_a|V\rangle_b+|V\rangle_a|H\rangle_b\right)
\end{equation}
This is a hyper-entangled state since photons are momentum entangled due to inseparability of $\Phi_{12}(r_{a};r_{b})$ and polarisation entangled due to inseparability of  $\frac{1}{\sqrt{2}} (|H\rangle_a|V\rangle_b+|V\rangle_a|H\rangle_b)$, where each quantum entangled state is symmetric.  However, a combined quantum state is a product of each quantum entangled state. From here onwards, a photon in region-$a$ is labeled as photon-1 and a photon in region-$b$ is labeled as photon-2.

\section{Imaging with hyper-entangled photons}
 In a realistic experiment, a source emits entangled photons in particular directions. To incorporate this directional dependence, quantum state is multiplied by an additional directional function such that the highest probability of finding photons is around a point $o_{a}$ in  region-$a$ and $o_{b}$ in  region-$b$. Let $\psi_{a}(x_{1},y_{1}, z_{1})$ and $\psi_{b}(x_{2},y_{2}, z_{2})$ are the envelops of wavefunctions for a photon around $o_{a}$ and a photon around $o_{b}$. From these considerations, a total symmetric quantum state of photons can be written as
 \begin{equation}\label{eq6}
 |\Psi\rangle_{s}=\frac{1}{\sqrt{2}}[\psi_{a}(x_{1},y_{1}, z_{1})\psi_{b}(x_{2},y_{2}, z_{2})
 \pm \\ \psi_{a}(x_{2},y_{2}, z_{2})\psi_{b}(x_{1},y_{1}, z_{1})]a_{s}\Phi_{12}(r_{a};r_{b})\otimes\frac{1}{\sqrt{2}} \left(|H\rangle_1|V\rangle_2+e^{i\phi}|V\rangle_1|H\rangle_2\right)
 \end{equation}
Where the phase $\phi$ is not arbitrary, it is either zero or $\pi$. The total quantum state is symmetrised with a plus sign if $\phi=0$. For $\phi=\pi$, the polarisation entangled state is antisymmetric after the exchange of photons therefore, the quantum state in the external degrees of freedom has to be antisymmetric. Here, $a_{s}$ is a normalisation constant. In the experiment, a phase difference $\phi$ between $|H\rangle$ and $|V\rangle$ quantum states of a photon is introduced either by placing a half-wave plate in the path of a photon or by tilting the nonlinear crystal source of entangled photons. In this paper, an antisymmetric polarisation entanglement is produced by the source.
Since photons are propagating in different regions $a$ and $b$ such that $\psi_{a}(x_{1},y_{1}, z_{1})$ and $\psi_{b}(x_{2},y_{2}, z_{2})$ are non-overlapping therefore, a second term $\psi_{a}(x_{2},y_{2}, z_{2})\psi_{b}(x_{1},y_{1}, z_{1})a_{s}\Phi_{21}(r_{a};r_{b})$, in Eq.~\ref{eq6} is negligible.

To form an image with a hyper-entangled state, consider photon-$2$ in region-$b$ is passed through a transparent polarisation sensitive phase pattern oriented perpendicular to $z$-axis as shown in Fig.~\ref{fig1}. The pattern introduces a position-dependent phase difference between the horizontal and the vertical polarisation of a photon passing through it such that $|H\rangle_{2}\rightarrow  e^{i\phi(x_{2}, y_{2})}|H\rangle_{2} $ and $|V\rangle_{2}\rightarrow |V\rangle_{2} $.  In the experiment, a phase difference $\phi(x_{2},y_{2})$ is either zero or $\pi$.  Consider photon-2 of a hyper-entangled state, with an antisymmetric polarisation entangled state $|\Psi^-\rangle_{p}=\frac{1}{\sqrt{2}}\left(|H\rangle_1|V\rangle_2-|V\rangle_1|H\rangle_2\right)$, is incident on the pattern. A polarisation dependent phase is imprinted on photon-2 in region-$b$. Assume that the pattern is located at $z=d_{2}$ in region-$b$. An arbitrary location of photon-2 on the pattern just after the phase imprint is represented by  coordinates ($x_{2}$, $y_{2}$) in a $x$-$y$ plane. An arbitrary location of photon-1 in a plane oriented perpendicular to $z$-axis at $z=-d_{1}$ in region-$a$ is represented by  coordinates ($x_{1}$, $y_{1}$).
Therefore, after the phase imprint and normalisation, the quantum state of photons is written as
 \begin{equation}\label{eq7}
 |\Psi\rangle_{I}=a_{s}\psi_{a}(x_{1},y_{1}, -d_{1})\psi_{b}(x_{2},y_{2}, d_{2})\Phi_{12}(r_{a};r_{b})
 \otimes\frac{1}{\sqrt{2}} \left(|H\rangle_1|V\rangle_2-e^{i\phi (x_{2}, y_{2})}|V\rangle_1|H\rangle_2\right)
 \end{equation}
A transparent polarisation sensitive phase pattern is imprinted in the phase of polarisation entangled state and a spread of a photon on the pattern is due to the momentum entangled part of the hyper-entangled state. Photons are also spatially correlated due to their spatial entanglement in  position space. The position dependent polarisation entangled state of photons is $|\Psi^-\rangle_{p}$ for $\phi (x_{2}, y_{2})=0$ and $|\Psi^+\rangle_{p}=\frac{1}{\sqrt{2}}\left(|H\rangle_1|V\rangle_2+|V\rangle_1|H\rangle_2\right)$ for $\phi (x_{2}, y_{2})=\pi$. The location and polarisation information of each photon is carried by the other photon because of  their quantum entanglement.

After the imprint of a pattern, the position coordinates ($x_{2}$, $y_{2}$) at $z=d_{2}$ of photon-$2$ are measured by imaging with a lens on a single photon sensitive camera after its polarisation selection by polariser-$2$ as shown in  Fig.~\ref{fig1}.  The camera registers photon-$2$ only if photon-$1$ is detected by a single photon detector after passing through polariser-$1$. Consider a lens $L_{o}$ is not placed but a narrow aperture single photon detector is placed at a location $(x_{o1},y_{o1}, z=-d_{1})$ and orientation of pass axis of polariser-$1$ is vertical and of polariser-$2$ is horizontal. This measurement setting corresponds to a coincidence detection when photon-1 is detected at $(x_{o1},y_{o1}, z=-d_{1})$ in the vertical polarisation and photon-$2$ is detected in the horizontal polarisation state on the camera. The same experiment is repeated for a chosen measurement setting many times, a location of photon-$2$ on camera varies each time and gradually an image is formed on the camera.
A corresponding probability of coincidence detection is written as
$P_{V_{1},H_{2}}(x_{2}, y_{2})=\frac{1}{2}| a_{s}\psi_{a}(x_{o1},y_{o1}, -d_{1})\psi_{b}(x_{2},y_{2}, d_{2})\Phi_{12}(r_{a};r_{b})(-e^{i\phi (x_{2}, y_{2})})|^{2}$. Where  subscripts of $P_{V_{1},H_{2}}$ denote orientation of pass-axes of polariser-$1$ and polariser-$2$. It is clear that a phase information is lost for this polarisation setting. Similarly, a polarisation sensitive phase information cannot be recovered for other combination of orientations of polarisers along horizontal and vertical directions.

Let the orientation of pass axes of polarisers is now changed to perform polarisation measurements in the diagonal basis that is $|d^{+}\rangle_{j}=\frac{|H\rangle_{j}+|V\rangle_{j}}{\sqrt{2}}$ and $|d^{-}\rangle_{j}=\frac{|H\rangle_{j}-|V\rangle_{j}}{\sqrt{2}}$, where a photon label $j$ is $1$ or $2$. If both polarisers are oriented to pass quantum states $|d^{-}\rangle_{j}$ and photon-$1$ is detected at a location  $(x_{o1},y_{o1}, z=-d_{1})$, then the corresponding probability of coincidence detection is
 \begin{equation}\label{eq8}
P_{d^{-}_{1},d^{-}_{2}}(x_{2},y_{2})=\frac{1}{4}| a_{s}\psi_{a}(x_{o1},y_{o1}, -d_{1})\psi_{b}(x_{2},y_{2}, d_{2}) \Phi_{12}(r_{a};r_{b})|^{2}
 (1-\cos\phi(x_{2},y_{2}))
 \end{equation}
which contains a phase information of the pattern. Since $\phi(x_{2},y_{2})$ is either $\pi$ or zero therefore, $P_{d^{-}_{1},d^{-}_{2}}(x_{2},y_{2})$ is a two level image. Where an image of the transparent polarisation sensitive phase pattern is obtained by measuring coincidence detection probability.
 If on the other hand, polariser-$1$ is aligned to pass $|d^{+}\rangle$ and polariser-$2$ is aligned to pass $|d^{-}\rangle$ then the corresponding probability of coincidence detection is written as
 \begin{equation}\label{eq9}
P_{d^{+}_{1},d^{-}_{2}}(x_{2},y_{2})=\frac{1}{4}| a_{s}\psi_{a}(x_{o1},y_{o1}, -d_{1})\psi_{b}(x_{2},y_{2}, d_{2}) \Phi_{12}(r_{a};r_{b})|^{2}
(1+\cos\phi(x_{2},y_{2}))
 \end{equation}
This image is exactly inverted as compared to  Eq.~\ref{eq8}, where the  maximum and the  minimum levels of the image are interchanged.

To make a measurement on photon-$1$ in a quantum superposition basis of position $(x_{1}, y_{1})$, a convex lens $L_{o}$ is placed at $z=-d_{1}$ and photon-$1$ is detected close to its focal point by a narrow aperture single photon detector placed on $z$-axis. The lens transforms a particular quantum superposition of position (or an incident momentum state) in a plane at $z=-d_{1}$ to a single point in its focal plane. Therefore, probability of coincidence detection of photons for two different orientation $(d^{+}_{1},d^{-}_{2})$ and $(d^{-}_{1},d^{-}_{2})$ of polarisers is succinctly written as
 \begin{equation}\label{eq10}
P'_{d^{\pm}_{1},d^{-}_{2}}(x_{2},y_{2})=\frac{1}{4} |\int^{\infty}_{-\infty}\int^{\infty}_{-\infty} a_{s}\psi_{a}(x_{1},y_{1}, -d_{1}) \psi_{b}(x_{2},y_{2}, d_{2})\Phi_{12}(r_{a};r_{b}) \mathrm{d}x_{1} \mathrm{d}y_{1}|^{2}
  (1\pm\cos\phi(x_{2},y_{2}))
 \end{equation}
 It is evident that opposite level images are formed when orientation of polarisers is changed from $(d^{-}_{1},d^{-}_{2})$ to $(d^{+}_{1},d^{-}_{2})$. However, no image is formed if only photon-2 is detected without considering the measurement outcome of photon-$1$ or if photon-$1$ is not measured at all.

\section{Experiment and discussion}
\begin{figure}
\centering
\includegraphics[scale=1.25]{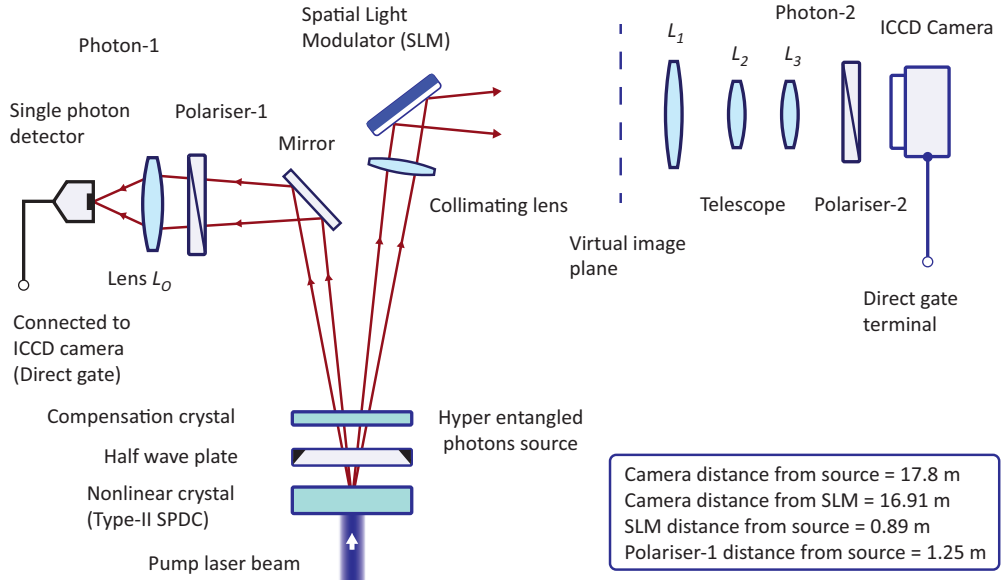}
\caption{\label{fig2} \emph{Main schematic diagram of the experiment. Hyper-entangled photons are produced by a BBO crystal. A polarisation sensitive phase pattern is produced by an SLM. The pattern is imaged in free space by ICCD camera located at 16.91~m from the pattern by a lens combination in a telescopic configuration.  ICCD camera is activated by a measurement outcome event signal of photon-1 to register photon-2 to form a quantum image.}}
\end{figure}

A schematic diagram of the experimental setup is shown in Fig.~\ref{fig2}. Hyper-entangled photons are produced by type-II spontaneous parametric down conversion (SPDC) \cite{mon1, mon2, souto, howell, stch, mandel_th} in a beta-barium-borate (BBO) nonlinear crystal. A linearly polarised pump laser beam of wavelength 405~nm and narrow beam diameter is passed through a BBO crystal. Hyper-entangled photons are produced at wavelength  810~nm in a first crystal, where a second crystal is placed to compensate transverse and longitudinal walk-offs. A polarisation sensitive phase pattern is produced by a reflection type spatial light modulator (SLM). After the collimation, photon-$2$ is reflected from the spatial light modulator and a position-dependent phase is imprinted such that $|H\rangle_{2}\rightarrow e^{i\phi(x_{2},y_{2})}|H\rangle_{2}$ and $|V\rangle_{2}\rightarrow|V\rangle_{2}$. Photon-$2$ is imaged on an intensified-charge-coupled-device (ICCD) camera, after passing through polariser-$2$, with a three lens ($L_{1}$, $L_{2}$ and $L_{3}$) telescopic configuration. The telescope is aligned such that the object plane coincides with the reflecting surface of the spatial light modulator. In this way, a location of photon-2 immediately after the phase imprint is measured by imaging it on the camera. The focal length of lens $L_{1}$ of diameter 5~cm is 40~cm. The focal length of lenses $L_{2}$ and $L_{3}$ are 10~cm and 100~cm, respectively. Photon-$1$ is passed through polariser-$1$ to perform a polarisation measurement and a lens $L_{o}$ to perform a measurement in the quantum superposition basis of position after detection by a narrow area single photon detector of very low dark counts. These are joint measurements. Pass-axis of polariser-$1$ is subtending an angle $\delta_{1}$ and of polariser-$2$ is subtending an angle $\delta_{2}$ with the horizontal axis.  These angles can be varied independently according to a measurement setting. An electrical signal produced by a single photon detector corresponds to a measurement outcome for a chosen  measurement setting for photon-$1$. This signal is connected to a direct gate terminal of the ICCD camera. The direct gate terminal activates an electronic shutter of the camera that allows detection signal of photon-$2$ produced by the imaging sensor of camera to reach the charge-coupled-device (CCD) sensor of the camera after amplification. This process efficiently detects a coincidence event of photons with a 20~ns insertion delay. The distance of the imaging sensor surface of the ICCD camera from the spatial light modulator is 16.91~m. The spatial light modulator is placed at a distance 89~cm from the first BBO crystal. A first lens $L_{1}$ of the telescope is placed at a distance 15.24~m from the spatial light modulator surface. This lens produces an inverted real image of the spatial light modulator surface, which acts as a real image for a second lens $L_{2}$. A second lens is placed such that it produces a magnified virtual image in a virtual image plane as shown in Fig.~\ref{fig2}. A third lens $L_{3}$ of focal length $f$~=~100~cm is positioned such that the photon detection plane of the ICCD camera and a virtual image plane are positioned at $+2f$ and $-2f$, respectively from lens $L_{3}$. In this $2f$-$2f$ imaging configuration, an image can be fine-tuned on the ICCD camera by displacing the second lens $L_{2}$ only and keeping other lenses stationary. The image demagnification of  the telescope is 0.52 and its spatial resolution at the spatial light modulator surface is about $\sim$~0.3~mm. In the experiment, a structure size of the pattern is 1~mm, which is imaged by the ICCD camera from a distance 16.91~m.
\begin{figure}[ht]
\centering
\includegraphics[scale=0.35]{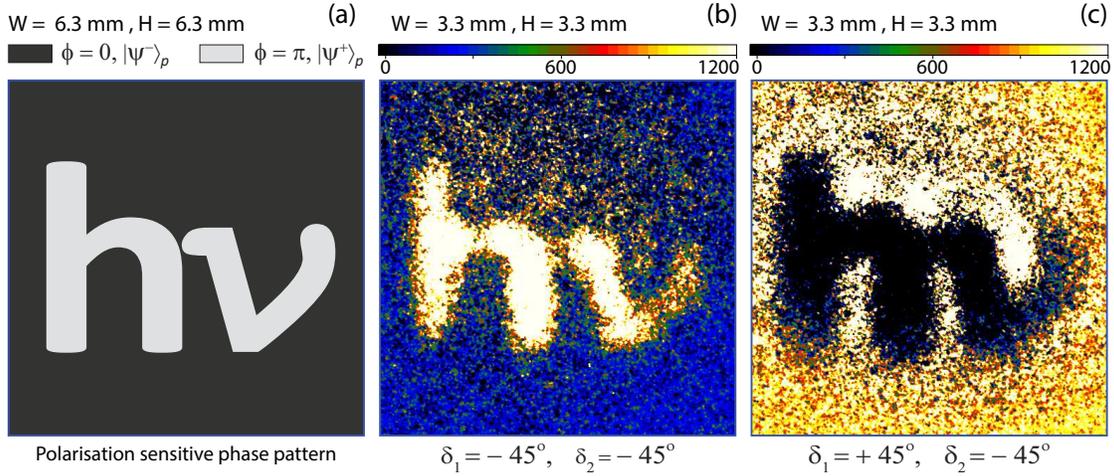}
\caption{\label{fig3} \emph{(a) A transparent polarisation sensitive phase pattern, a darker region represents $\phi(x_{2},y_{2})=0$ and a lighter region represents $\phi(x_{2},y_{2})=\pi$. (b) An image captured by ICCD camera, when measured polarisation states of photon-1 and photon-2 are $|d^{-}\rangle_{1}$ and $|d^{-}\rangle_{2}$. (c) An image with inverted levels, when measured polarisation state of photon-1 is  $|d^{+}\rangle_{1}$ and of photon-2 is $|d^{-}\rangle_{2}$. Each image is captured for ten minutes time of exposure.}}
\end{figure}

\begin{figure}[ht]
\centering
\includegraphics[scale=0.3]{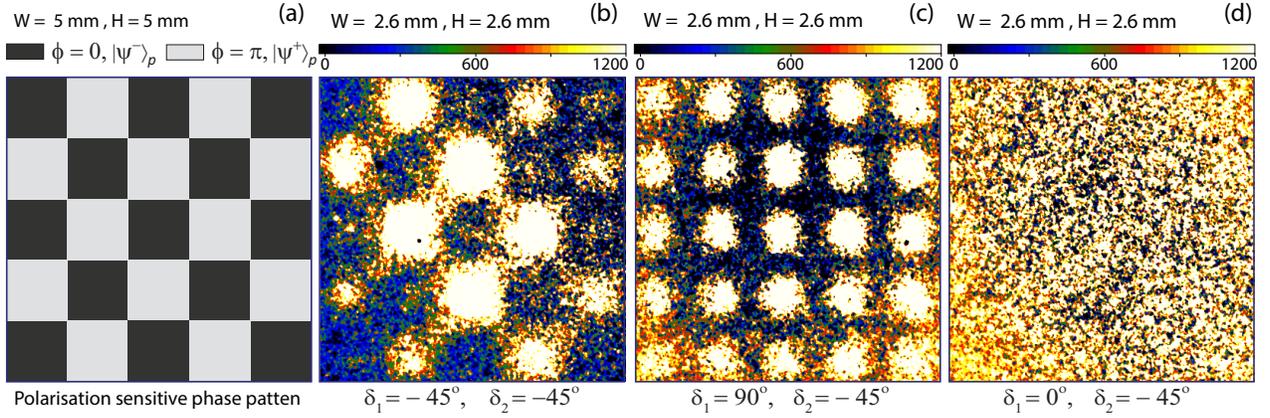}
\caption{\label{fig4} \emph{ (a) A transparent polarisation sensitive phase pattern with $1~mm\times 1~mm$ size of each square, a darker region represents $\phi(x_{2},y_{2})=0$ and a lighter region represents $\phi(x_{2},y_{2})=\pi$. (b) An image of the pattern captured by ICCD camera, when measured polarisation states of photon-1 and photon-2 are $|d^{-}\rangle_{1}$ and $|d^{-}\rangle_{2}$. (c) When measured polarisation state of photon-1 is $|V\rangle_{1}$ and of photon-2 is $|d^{-}\rangle_{2}$. In this case, only the edges of squares are absent due to the diffraction limit of telescope. (d) No image is formed when measured polarisation state of photon-1 is $|H\rangle_{1}$ and of photon-2 is $|d^{-}\rangle_{2}$.}}
\end{figure}
Consider, photon-$j$ is detected after passing through a polariser-$j$ with its pass axis inclined at an angle $\delta_{j}$ \emph{w.r.t.} horizontal axis. Its measured state of polarisation is, $|H\rangle_{j}$ for $\delta_{j}=0$, $|V\rangle_{j}$ for $\delta_{j}=90^{o}$, $|d^{+}\rangle_{j}$ for $\delta_{j}=45^{o}$ and $|d^{-}\rangle_{j}$ for $\delta_{j}=-45^{o}$. Polarisation state of photon-$2$ is measured once it is detected by ICCD camera at any location after passing through polariser-$2$. If a single photon detector detects photon-$1$ after passing through a polariser-$1$ but without a lens $L_{o}$, then it corresponds to a position measurement of photon-1 $(x_{01}, y_{01})$ in a plane at $z=-d_{1}$ along with an outcome of measured state of polarisation of photon-$1$. On the other hand, if a single photon detector detects a photon-$1$ after passing through a polariser and a lens $L_{o}$, then it corresponds to a measurement in the quantum superposition basis of position along with an outcome of a measured state of polarisation of photon-$1$. For this measurement setting, a coincidence photon detection probability is given by Eq.~\ref{eq10}.

A transparent polarisation sensitive phase pattern shown in Fig.~\ref{fig3}(a) is displayed on the spatial light modulator, where a position-dependent grey level of the pattern determines a phase shift, which is either zero or $\pi$. Lighter grey regions represent a $\pi$ phase shift and since a source produces an antisymmetric polarisation entangled state therefore, polarisation entanglement of  photons, after passing photon-$2$ through this region, transformed to $|\Psi^{+}\rangle_{p}=\frac{1}{\sqrt{2}} (|H\rangle_1|V\rangle_2+|V\rangle_1|H\rangle_2)$. Darker grey regions represents a zero phase shift and polarisation entangled state of photons remains same as produced by the source that is $|\Psi^{-}\rangle_{p}=\frac{1}{\sqrt{2}} (|H\rangle_1|V\rangle_2-|V\rangle_1|H\rangle_2)$. Therefore, the information of a two-photon quantum image is contained by the momentum and the polarisation entanglement parts of $|\Psi\rangle_{I}$. In the first experiment, to produced a quantum image, both polarisers are aligned at $\delta_{1}=-45^{o}$ and $\delta_{2}=-45^{o}$ to detect a same state of polarisation of photons in the diagonal basis that is $|d^{-}\rangle_{1}$ of photon-$1$  and $|d^{-}\rangle_{2}$ of photon-$2$. Photon-1 is also measured in the quantum superposition basis of position. The experiment is repeated for this setting and an image, in the form of gated photon counts of the pattern, is accumulated for 10~minutes exposure of the ICCD camera. A background image is captured by displaying a uniform grey level corresponding to $\phi=0$ on the spatial light modulator with the same experimental setting. A coincidence photon image is constructed by subtracting a corresponding background image as shown in  Fig.~\ref{fig3}(b).  The ICCD camera also captures accidental counts in the coincidence window. These accidental counts are originated from the scattered photons and camera noise. These counts accumulate with the actual coincidence photon counts. A typical ratio of a coincidence to accidental counts is 0.35-0.45. This is why a background correction is necessary. The noise of accidental photon counts in the background image, which is defined as a ratio of standard deviation to an average of counts for twenty-five hundred pixels is typically 0.12. All images shown in this paper are background corrected. Each image on the ICCD camera is reduced by 0.52 image demagnification.  In a similar experiment, polariser-$1$ is aligned at $\delta_{1}=+45^{o}$ to detect photon-$1$ in $|d^{+}\rangle_{1}$  and polariser-2 is kept at $\delta_{2}=-45^{o}$ to detect photon in $|d^{-}\rangle_{2}$ .  The coincidence image photon counts accumulated by  the ICCD camera are shown in Fig.~\ref{fig3}(c), which corresponds to an image with inverted levels as compared to an image shown in Fig.~\ref{fig3}(b). This result is in agreement with Eq.~\ref{eq10} that shows an inversion of levels of images in the diagonal basis. These images are two-level images corresponding to the two-level phase shift of the pattern.

\begin{figure*}[ht]
\centering
\includegraphics[scale=0.35]{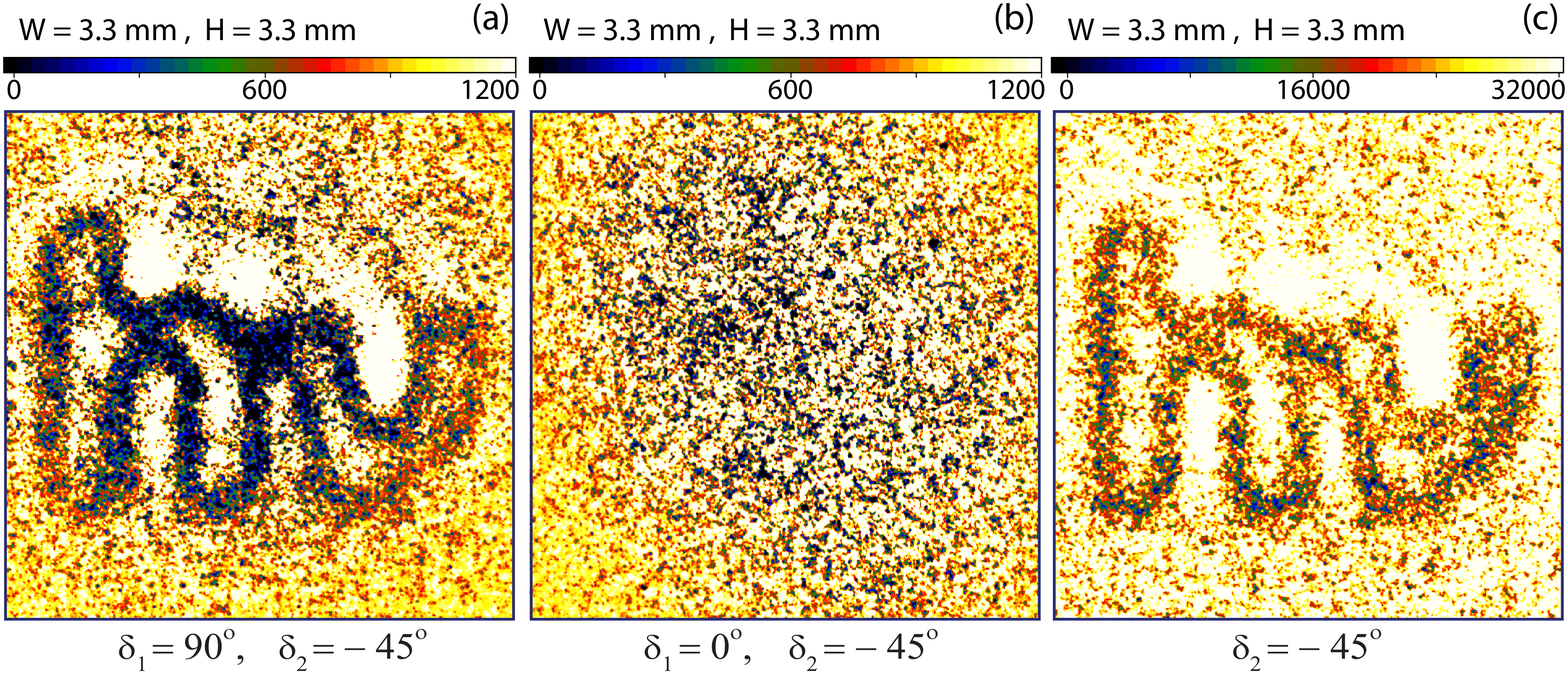}
\caption{\label{fig5} \emph{(a) Edges of a pattern shown in  Fig.~\ref{fig3}(a) appear when measured polarisation state of photon-1 is $|V\rangle_{1}$  and of photon-2 is $|d^{-}\rangle_{2}$. (b) No image is formed when measured polarisation state of photon-1 is $|H\rangle_{1}$ and of photon-2 is $|d^{-}\rangle_{2}$. (c) A single photon image of low depth
captured by ICCD camera. Edges of the pattern appear as low coincidence photon counts darker regions due to the resolution limit of the telescope.}}
\end{figure*}

In another experiment, a different transparent polarisation sensitive phase pattern as shown in Fig.~\ref{fig4}(a) is displayed on the spatial light modulator. Where the gray levels corresponding to the position-dependent phase shifts are the same as in the previous pattern. A background-corrected image of this pattern shown in Fig.~\ref{fig4}(b) is accumulated with the ICCD camera for 10~minutes time of exposure in the diagonal basis, for $\delta_{1}=\delta_{2}=-45^{o}$ and with a lens $L_{o}$. This image is then compared with another background-corrected image captured with a different setting of polariser-$1$, $\delta_{1}=+90^{o}$ as shown in Fig.~\ref{fig4}(c) keeping all other settings the same.  According to the theoretical analysis shown in the previous section, the coincidence probability corresponding to a measurement of $|H\rangle$ or $|V\rangle$ state of photon-$1$ does not contain the phase information of the pattern therefore, no image should be formed. A measurement of photon-$1$ polarisation state in $|V\rangle_{1}$ collapses photon-$2$ polarisation state onto $|H\rangle_{2}$ and this polarisation state is transformed by the spatial light modulator such that $|H\rangle_{2}\rightarrow e^{i\phi(x_{2},y_{2})}|H\rangle_{2}$. This is a pure phase pattern and it diffracts photons. However, a telescope is located at a far distance 15.24~m  from the pattern and its spatial resolution at this distance is about $\sim$~0.3~mm, where the size of each square in the pattern is 1~mm$\times$1~mm. Due to the finite lens diameter, the telescope is not capturing all the diffracted photons carrying information of sharp edges and it leads to the fundamental diffraction limit of resolution. Thus, the edges or phase boundaries in the pattern lead to low two-photon coincidence counts. Therefore, the edges appear as low-level regions in the image. In this case, if all diffracted photons are also captured then no image of the pattern will result. This effect is exactly matching for all the sharp edges by a comparison of an image shown in Fig.~\ref{fig4}(c) with a corresponding image shown in Fig.~\ref{fig4}(b). Another image is captured for setting, $\delta_{1}=0$ to detect photon-$1$ in $|H\rangle_{1}$ keeping all other settings the same as for the image shown in Fig.~\ref{fig4}(b). Since this setting collapses the polarisation state of photon-$2$ to $|V\rangle_{2}$ and this quantum state remains unchanged by the spatial light modulator therefore, no pattern is formed on ICCD camera as shown in Fig.~\ref{fig4}(d).
A similar experiment is repeated for $\delta_{1}=90^{o}$  to detect $|V\rangle_{1}$ and $\delta_{2}=-45^{o}$ to detect $|d^{-}\rangle_{2}$ for a pattern shown in Fig.~\ref{fig3}(a). A corresponding background-corrected image is shown in Fig.~\ref{fig5}(a), where the sharp edges appear due to the diffraction limit of the telescope. As explained, for $\delta_{1}=0$ and $\delta_{2}=-45^{o}$, no pattern is formed as shown in Fig.~\ref{fig5}(b). A single photon background-corrected image of the same pattern is shown in Fig.~\ref{fig5}(c). In this case, photon-1 is not measured and the ICCD camera shutter is kept open for continuous exposure of photon-2 for ten minutes. This image has low depth and only the sharp edges appear because of the diffraction limit of the telescope.

\section{Conclusion}
This paper presents quantum imaging experiments of transparent polarisation sensitive phase patterns with hyper-entangled photons. The paper begins with the main concept of the experiment and then it is shown, how a hyper-entangled state arises from the indistinguishability of identical photons. Furthermore, a detailed theoretical analysis of quantum imaging is presented. In the experiment, hyper-entangled photons are produced by a type-II SPDC source and each polarisation sensitive phase pattern is imaged from a distance 16.91~m with the ICCD camera. In this type of coincidence imaging, polarisation entanglement and momentum entanglement of photons are involved. Images of patterns are constructed after background correction by accumulating coincidence photons on the ICCD camera for different measurement settings for both photons. In a diagonal basis, image levels are inverted if a measured outcome of polarisation state of photon-$1$ is changed from  $|d^{-}\rangle_{1}$ to $|d^{+}\rangle_{1}$. Edges of the pattern are not captured because of the diffraction limit of the telescope. Eventually, low-level regions corresponding to edges appear in a two-photon captured image. Which is experimentally shown in detail. If photon-$1$ is measured in a quantum state $|H\rangle_{1}$ then photon-$2$ collapses to a quantum state $|V\rangle_{2}$, which is unaffected by the spatial light modulator and it leads to no pattern in the image. A single photon accumulation of photon-$2$ on the camera by continuous exposure without considering photon-$1$ produces a shallow image due to the diffraction limit of the telescope, which contains partial information of the pattern.

\section*{Methods}
\subsection{Experimental setup}
Hyper-entangled photons are produced by a type-II SPDC process in a first BBO crystal, which is a negative uniaxial crystal. A focused pump laser beam of extraordinary polarisation is incident on the crystal. Crystal is tuned such that the emission cones of ordinary and extraordinary polarised photons are intersecting. Hyper-entanglement is produced in the intersection regions. Down converted photons at wavelength 810~nm are passed through a half-wave plate to interchange their linear polarisation. A second crystal of half thickness is placed parallel to the first crystal with a same orientation of its optic axis as the first crystal. This crystal compensates for transverse and longitudinal walk-offs of down converted photons. The source is aligned to produce an antisymmetric polarisation entangled state $|\Psi^-\rangle_{p}$ and a momentum entangled state. In a position basis, the momentum entangled photons are spatially entangled. Momentum entanglement is a result of the linear momentum conservation of photons. Photons are polarisation entangled and momentum entangled separately. Photon-$1$ and photon-$2$ of an entangled pair are passed through two polarisers aligned with their pass axes subtending angles $\delta_{1}$ and $\delta_{2}$ with the horizontal direction. The photons transmitted by the polarisers are passed through bandpass filters of peak transmission at 810~nm. To check polarisation entanglement the photons are then detected by single photon detectors where single photon counts of each photon detector and coincidence photon counts are measured by a coincidence event counting module. However, to image the pattern the experimental setup is shown in Fig.~\ref{fig2}.

\subsection{Violation of CHSH inequality}

Einstein's locality condition strictly forbids the faster than light influence. Consider two polarisation entangled photons separated by a large distance. Any measurement performed on one photon should not affect the measurement outcome of the other photon immidiately if the locality condition is true. Consider two different observables $A_{1}$ and $A'_{1}$ of photon-$1$ where any one of them is measured for a photon pair. Similarly, any one of the two different observables $B_{2}$ and $B'_{2}$ of photon-$2$ is measured for a photon pair. A measured observable outcome corresponds to a polarisation measurement of a photon. A measurement outcome corresponds to one of the two orthogonal polarisations with measured value $+1$ for one polarisation and  $-1$ for the other. In the local realistic model, the values of all such observables are well-defined prior to any measurement. Consider a correlation function where each observable can have a value +1 or -1, therefore
\begin{equation}
\label{eqm1}
  C_{l}= (A_{1}+A'_{1})B'_{2}+(A_{1}-A'_{1})B_{2}=\pm2
\end{equation}
its value cannot exceed the bound $|\langle C_{l}\rangle|\leq2$.
However, quantum mechanical observables cannot always have a well-defined value and incompatible observables cannot be measured simultaneously. An observable value is obtained by the measurement as a possible outcome that depends on the quantum state and observable being measured. Quantum mechanics violates local realistic prediction expressed in quantum observables such that
\begin{eqnarray}
\label{eqm2}
S=\langle (\hat{A}_{1}+\hat{A}'_{1})\hat{B}'_{2}+(\hat{A}_{1}-\hat{A}'_{1})\hat{B}_{2} \rangle
=\langle \hat{A}_{1}\hat{B}_{2}\rangle-\langle\hat{A}'_{1}\hat{B}_{2}\rangle+\langle\hat{A}_{1}\hat{B}'_{2}\rangle+\langle\hat{A}'_{1}\hat{B}'_{2}\rangle
\end{eqnarray}
Therefore, the following Clauser, Horne, Shimony and Holt (CHSH) inequality is violated by quantum mechanics \cite{chsh_th}
\begin{equation}
0\leq|S|\leq 2
\end{equation}

\begin{figure}
\centering
\includegraphics[scale=0.6]{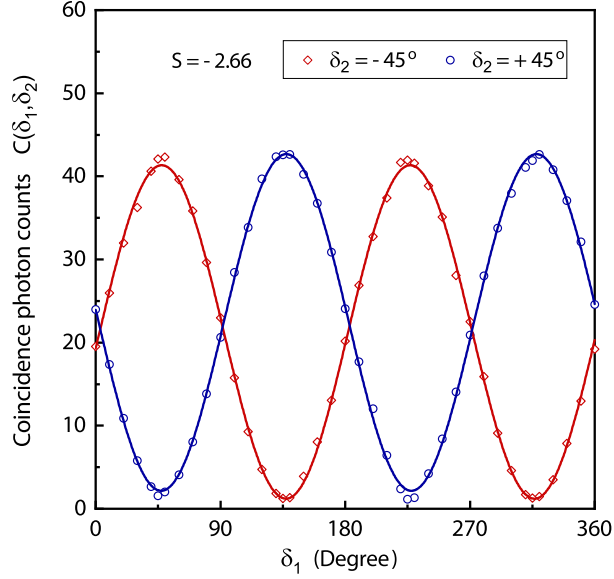}
\caption{\label{mtfig_1} \emph{Coincidence photon counts measurements for $|\Psi^{-}\rangle_{p}$ at different orientations of pass axes of polarisers. Solid line represents a fit of  $\sin^{2}(\delta_{1}-\delta_{2})$ corresponding to the coincidence detection probability.}}
\end{figure}

Expectation value of each term of product of observables in Eq.~\ref{eqm2} can be evaluated from coincidence photon counts $C(\delta_{1}, \delta_{2})$ measured for a corresponding orientation of pass axes of polarisers and from the coincidence photon counts $C(\delta^{\perp}_{1}, \delta^{\perp}_{2})$  where $\delta^{\perp}_{j}=\delta_{j}+90^{o}$ \cite{chsh}.
The CHSH inequality is violated for the polarisation entanglement. A plot of coincidence photon counts measurements is shown in Fig.~\ref{mtfig_1} for different orientations of polarisers. The CHSH correlation parameter $S$, Eq.~\ref{eqm2}, is expressed in the form of angle of pass axes of polarisers as
\begin{equation}
\label{eqm3}
  S=E(\delta_{1},\delta_{2})-E(\delta'_{1},\delta_{2})+E(\delta_{1},\delta'_{2})+E(\delta'_{1},\delta'_{2})
\end{equation}
which is evaluated from the coincidence photon measurements. Where $E(\delta_{1},\delta_{2})$ represents an expectation value of the joint observable measurements performed on both photons, which is written as
\begin{equation}
\label{eqm4}
  E(\delta_{1},\delta_{2})=\frac{C(\delta_{1}, \delta_{2})+C(\delta^{\perp}_{1}, \delta^{\perp}_{2})-C(\delta_{1}, \delta^{\perp}_{2})-C(\delta^{\perp}_{1}, \delta_{2})}{C(\delta_{1}, \delta_{2})+C(\delta^{\perp}_{1}, \delta^{\perp}_{2})+C(\delta_{1}, \delta^{\perp}_{2})+C(\delta^{\perp}_{1}, \delta_{2})}
\end{equation}

A measured value of parameter $S=-2.66$ where $0\leq|S|\leq 2$ for a local realistic model, thus CHSH inequality is violated. The CHSH inequality is also violated by taking coincidence photon measurements of observables with the ICCD camera gated by the single photon detector signal.

\subsection{Spatial correlations}
In the regions of the intersection of cones, the down converted photons are diverging and are spatially correlated in the transverse planes as a consequence of their spatial entanglement. These correlations are a result of momentum entanglement of photons which is represented by $\Phi_{12}(r_{a};r_{b})$ in Eq.~\ref{eq5}. Spatial entanglement correlates locations of photons. It is due to the spread of photons wavefunction the whole pattern is exposed by a single photon. Once a location of a photon is determined on the pattern its information is shared with the other photon due to their spatial correlations thus each photon carries the spatial location information of the other photon. Position correlations are measured in two planes oriented perpendicular to the mean direction of propagation of each photon.
A single-slit of width 0.8~mm is placed at a distance $d_{1}=1.247$~m, immediately in front of a convex lens $L_{o}$, where the lens is placed in front of a narrow area single photon detector such that the detector is almost at its focal point. The length of the slit is very large as compared to its width. Polariser-$1$ and polariser-$2$  are removed from paths of photons and the remaining setup is the same as shown in Fig.~\ref{fig2}.
\begin{figure*}
\centering
\includegraphics[scale=0.60]{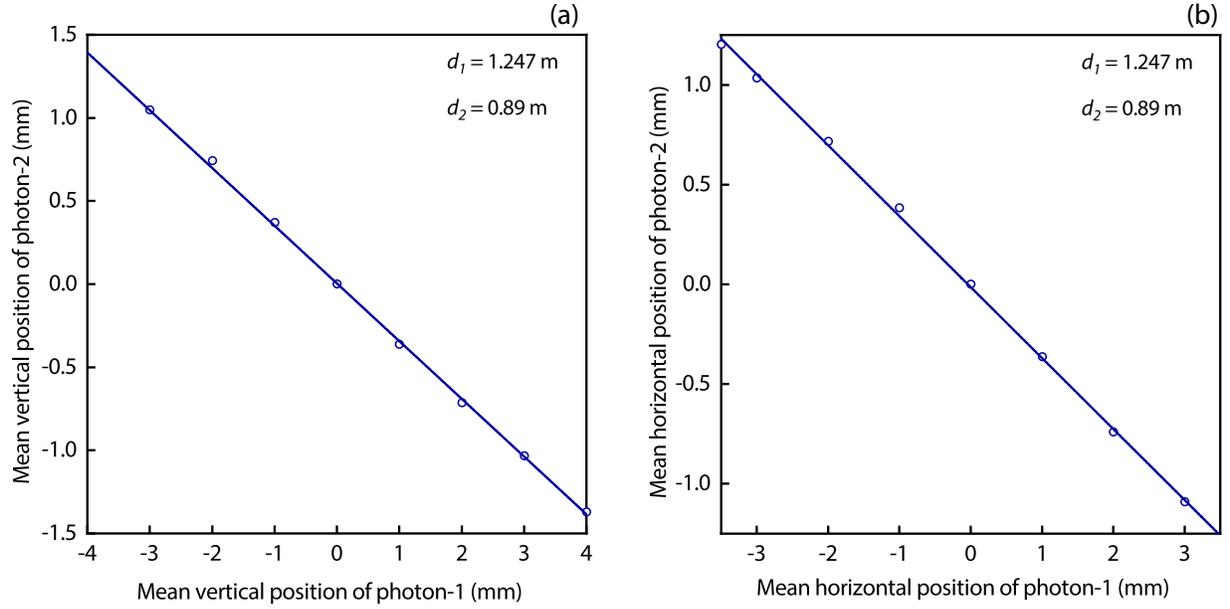}
\caption{\label{mtfig_2} \emph{ (a) Vertical position measurements of photons.  Each vertical position of photon-$1$ is measured by passing it through a horizontally aligned single narrow slit placed in front of a lens $L_{o}$. A corresponding position of photon-$2$ is measured by the ICCD camera in the plane of the spatial light modulator. Each data circle represents  mean vertical positions of photons corresponding to a particular position of a single slit. A straight line is a linear fit which shows that vertical positions of photons are equal and opposite at equal distances from the BBO crystal. (b)  Horizontal position measurements of photons when narrow single slit is aligned vertically and displaced horizontally. A straight line is a linear fit which shows that horizontal positions of photons are equal and opposite at equal distances from the BBO crystal.}}
\end{figure*}

If the single-slit is oriented horizontally and photon-$1$ is passed through it and detected then a vertical position of photon-1 is measured with a precision equals to the slit width because photon-$1$ is detected only if it is transmitted by the slit. However, this measurement does not measure horizontal position because of large length of the slit in  the horizontal direction. Similarly, a corresponding mean position of photon-$2$ is measured on the ICCD camera, which corresponds to a measured mean position of photon-$2$ on the spatial light modulator placed at $d_{2}=0.89$~m reduced by 0.52 due to the image demagnification of the telescope. Mean vertical positions of photon-$1$ and photon-$2$ are measured after repeating the experiment for different vertical displacements of the single-slit. Correlation between the mean vertical positions of photon-$1$ and of photon-$2$ is shown in Fig.~\ref{mtfig_2}(a).
Similar measurements are performed when a single-slit is oriented vertically and displaced horizontally. This setting performs a horizontal position measurement of photon-$1$ with a resolution equals to the slit width. Mean horizontal position of photon-$1$ and of photon-$2$ are measured by repeating the experiment for each horizontal displacement of the single-slit. A corresponding correlation of mean positions of photons is shown in Fig.~\ref{mtfig_2}(b). The telescope demagnification is 0.52 and distances $d_{1}$ and $d_{2}$ are different in the position correlation plots. For $d_{1}=d_{2}$ and after taking into account the demagnification, the measured mean position of photon-$1$ is equal and opposite to the measured mean position of photon-$2$ in the transverse planes.

\subsection{Classical and quantum imaging}
Most of the imaging methods in everyday applications are based on classical imaging. These methods utilise properties of classical fields such as intensity, polarisation, phase, frequency and coherence. In the case of optical imaging,  classical fields are electromagnetic wave fields. If more than one electromagnetic wave fields are involved then a classical image can be constructed by correlating their classical observable properties. This type of imaging technique, which is known as classical coincidence imaging, has many applications in biomedical imaging and astronomy. A classical ghost imaging is also a coincidence imaging technique, where only one electromagnetic field interacts with the object and its intensity is measured position-wise while the other electromagnetic wave is detected by a bucket detector. Most of these imaging methods can image absorptive objects only. A pure phase object is transparent and cannot be imaged by such methods. However, a method of phase-contrast imaging, which transforms a phase variation to intensity variation across the object is used to image a pure phase object. A phase-contrast microscope can see transparent micro-organisms and transparent objects. Such a method can also image a Bose-Einstein condensate non-destructively. In addition, a transparent polarisation sensitive phase object can be imaged with a polarisation-contrast microscope  \cite{pci1, pci2,pci3}, which measures the polarisation shift caused by the object. However, quantum imaging methods are  based on the quantum mechanical properties of photons originating from their polarisation quantum states, their quantum entanglement and quantum coherence. Quantum entanglement enhanced resolution microscope can image microscopic structures beyond the diffraction limit \cite{enh1}.  Quantum image construction also depends on how measurements are performed on each photon since incompatible observables can not be measured simultaneously. In experiments presented in this paper, two joint measurements of compatible observables are performed on each photon since they are hyper entangled in two different degrees of freedom. The first measurement is polarisation measurement in a chosen polarisation basis and the second measurement corresponds to a measurement of the momentum of the non-interacting photon and position for the interacted photon. The individual photon carries no complete image information and this method can be useful to send images of the polarisation sensitive phase pattern securely and directly over a large distance. The experiment shown in this paper is the first demonstration of quantum imaging with hyper-entangled photons over a long path.

\subsection*{Acknowledgement}
Mandip Singh acknowledges research funding by the Department of Science and Technology, Quantum Enabled Science and Technology grant for project No. Q.101 of theme title ``Quantum Information Technologies with Photonic Devices", {\bf{DST/ICPS/QuST/Theme-1/2019 (General)}}.

\subsection*{Author contributions statement}
 MS conceptualised the idea and setup the experiment, both authors performed the experiment, MK took data, MS wrote the manuscript and supervised the project.

\bibliography{sample}




\end{document}